\documentclass[aps,prb,twocolumn,superscriptaddress,longbibliography]{revtex4-1}

\usepackage{amsmath}
\usepackage{graphicx}
\usepackage{bm}
\usepackage{amssymb}
\usepackage{hyperref}
\usepackage{xcolor}
\usepackage{amsmath}
\usepackage{amssymb}

\begin{document}
\title{Gate-tunable Lifshitz transition of Fermi arcs and its nonlocal transport signatures}

\author{Yue Zheng}
\affiliation{National Laboratory of Solid State Microstructures, School of Physics,
and Collaborative Innovation Center of Advanced Microstructures, Nanjing University, Nanjing 210093, China}

\author{Wei Chen}
\email{Corresponding author: pchenweis@gmail.com}
\affiliation{National Laboratory of Solid State Microstructures, School of Physics,
and Collaborative Innovation Center of Advanced Microstructures, Nanjing University, Nanjing 210093, China}

\author{Xiangang Wan}
\affiliation{National Laboratory of Solid State Microstructures, School of Physics,
and Collaborative Innovation Center of Advanced Microstructures, Nanjing University, Nanjing 210093, China}

\author{D. Y. Xing}
\affiliation{National Laboratory of Solid State Microstructures, School of Physics,
and Collaborative Innovation Center of Advanced Microstructures, Nanjing University, Nanjing 210093, China}

\begin{abstract}
One hallmark of the Weyl semimetal is the emergence of Fermi arcs (FAs) in
the surface Brillouin zone that connect the projected Weyl nodes of opposite chirality.
The unclosed FAs can give rise to various exotic effects that have attracted
tremendous research interest. The configurations of the FAs
are usually thought to be determined fully by the band topology of
the bulk states, which seems impossible to manipulate.
Here, we show that the FAs can be simply modified by a surface gate voltage.
Because the penetration length of the surface states depends on the in-plane momentum,
a surface gate voltage induces an effective energy dispersion.
As a result, a continuous deformation of the surface band can be
implemented by tuning the surface gate voltage.
In particular, as the saddle point of the surface band meets the Fermi energy,
the topological Lifshitz transition takes place for the FAs,
during which the Weyl nodes switch their partners connected by the FAs.
Accordingly, the magnetic Weyl orbits composed of the FAs on opposite
surfaces and chiral Landau bands inside the bulk
change its configurations. We show that such an effect
can be probed by the nonlocal transport measurements
in a magnetic field, in which the switch on and off
of the nonlocal conductance by the surface gate voltage signals
the Lifshitz transition. Our work opens a new route for manipulating the FAs
by surface gates and exploring novel transport phenomena associated with
the topological Lifshitz transition.

\end{abstract}
\maketitle

\section{introduction}
In the last two decades, the research on novel topological materials
has seen rapid progress, involving the discoveries of the
topological insulators~\cite{RevModPhys.83.1057,RevModPhys.82.3045,PhysRevLett.95.226801}
and topological semimetals~\cite{armitage2018weyl,hasan2017discovery,wan2011topological,zhang2017topological,yu2017topological,li2020fermi,PhysRevB.84.235126}.
The latter possess gapless energy spectra but
nontrivial band topology, which can give rise to
interesting effects stemming from both the bulk and surface states.
According to the features of the band degeneracies,
topological semimetals can be further classified into several types, such as
Weyl semimetal (WSM), Dirac semimetal~\cite{wan2011topological,armitage2018weyl,hasan2017discovery}
and nodal-line semimetal~\cite{PhysRevB.84.235126}.
As the counterparts of the massless Weyl and Dirac fermions in condensed matter physics,
the quasiparticles with linear dispersion in the WSM and Dirac semimetals
provide an interesting platform for exploring novel effects predicted by high-energy physics~\cite{xu2015discovery,
xu2015discovery2,xu2015experimental,xu2016observation,
huang2016spectroscopic,tamai2016fermi,deng2016experimental,yang2015weyl,jiang2017signature,belopolski2016discovery,
lv2015observation,murakami2007phase,
huang2015weyl,lv2015experimental,Adler69pr,bell1969pcac}.
These effects are manifested as anomalous transport and optical properties
which can be probed using a standard approach of condensed matter physics~\cite{Zyuzin12prb,jian2013topological,burkov2015chiral,Aji12prb,Son13prb,
Chernodub14prb,Ma15prb,Zhong16prl,Spivak16prb,hirschberger2016chiral,Wang16prb}.

The nontrivial band topology of the WSMs is embodied in the monopole charge (or Chern number of the Berry
curvature field) carried by the Weyl nodes. According to the no-go theorem \cite{nielsen1981absence,nielsen1983adler}, the Weyl nodes of
opposite chirality must appear in pairs. The manifestation
of the nontrivial band topology of the WSM is the
unclosed Fermi arcs (FAs) spanning between Weyl nodes
of opposite chirality projected into the surface Brillouin zone.
The emergence of the FAs is a unique property
of the WSMs, without any counterpart in high-energy physics,
which can not only serve as the hallmark of the WSMs
\cite{huang2015weyl,lv2015experimental,xu2015discovery,
xu2015discovery2,xu2015experimental,xu2016observation,
deng2016experimental,yang2015weyl,huang2016spectroscopic,
tamai2016fermi,jiang2017signature,belopolski2016discovery,
lv2015observation},
but also lead to several novel phenomena~\cite{armitage2018weyl,hasan2017discovery,wang2017quantum,lu2017quantum,
zheng2018quasiparticle,Huang15prx,shekhar2015extremely,du2016large,zhang2016signatures}.
Given that the FAs are the Fermi surface of the topological surface states,
one might think that all its properties, especially how they connect
pairs of Weyl nodes are completely determined by the band topology
of the bulk states through the bulk-boundary correspondence.
Therefore, it seems that the only way to modify the configurations
of the FAs is to change the bulk properties of the WSMs.

Interestingly, recent research progress shows that the
configurations of the FAs are quite sensitive to the details of the sample boundary~\cite{morali2019fermi,yang2019topological,Ekahana20prb},
which opens the possibility for manipulating the FAs through surface modifications.
In particular, the topological Lifshitz transition~\cite{lifshitz1960anomalies} of the FAs can be induced by
surface decoration~\cite{yang2019topological} or chemical potential modification~\cite{Ekahana20prb},
which changes the sizes and shapes of the FAs, and especially, the way they connect pairs of Weyl nodes.
The existing experiments show that FAs with different configurations
can be realized in different samples~\cite{morali2019fermi,yang2019topological,Ekahana20prb,chen2018proposal,model,PhysRevB.104.075420,PhysRevB.104.205412},
but whether it is possible to continuously modify the FAs in a given sample remains an open question.
It is of great interest to explore the possibility of manipulating
the FAs by external fields, in which both continuous deformation
and abrupt Lifshitz transition of the FAs can be achieved.

In this work, we show that the FAs of the WSMs can
be continuously tuned by a surface gate voltage.
Because the penetration length of the surface state depends on the in-plane momentum,
the surface gate voltage acts
unequally on these surface modes and
induces a momentum-dependent potential energy,
or effectively, an additional dispersion of the surface band~\cite{model}.
As a result, a continuous deformation of the surface band and so the FAs can be achieved
by simply tuning the gate voltage.
It is shown that the existence of the saddle point in the surface band
is responsible for the topological Lifshitz transition
of the FAs. Specifically, the transition takes place
when the saddle point coincides with the Fermi energy.
At the same time, the Weyl nodes switch their partners
that are connected by the FAs. A direct physical result
is that the magnetic Weyl orbits composed of the FAs on opposite
surfaces and the chiral Landau bands inside the bulk
change their configurations. We show that such an effect
can be probed by the nonlocal transport measurements
in a magnetic field, in which the nonlocal current
can be switched on and off by the gate voltage,
thus providing a clear signature of the gate-voltage induced Lifshitz transition.
In addition to the study on the effective model,
we also calculate the surface electrostatic potential induced by the gate voltage
in a specific WSM ZrTe using the first-principles calculations.
It shows that the gate voltage
can induce large potential energy which is sufficient for
driving the Lifshitz transition of FAs.
Our work not only uncovers the scenario of the Lifshitz transition of the FAs
but also paves the way for its continuous manipulation by a surface gate voltage.

The rest of this paper is organized as follows: in Sec.~\ref{c2}, we study
the Lifshitz transition of the FAs based on both the continuous and lattice models.
In Sec.~\ref{c3}, we study the transport signatures of the surface gate induced Lifshitz transition.
In Sec.~\ref{c4}, we discuss the experimental implementation of our proposal.
Finally, we give a brief summary in Sec.~\ref{c5}.

\section{Lifshitz transition of Fermi-arc induced by surface gate}\label{c2}

\begin{figure}
\center
\includegraphics[width=1.0\linewidth]{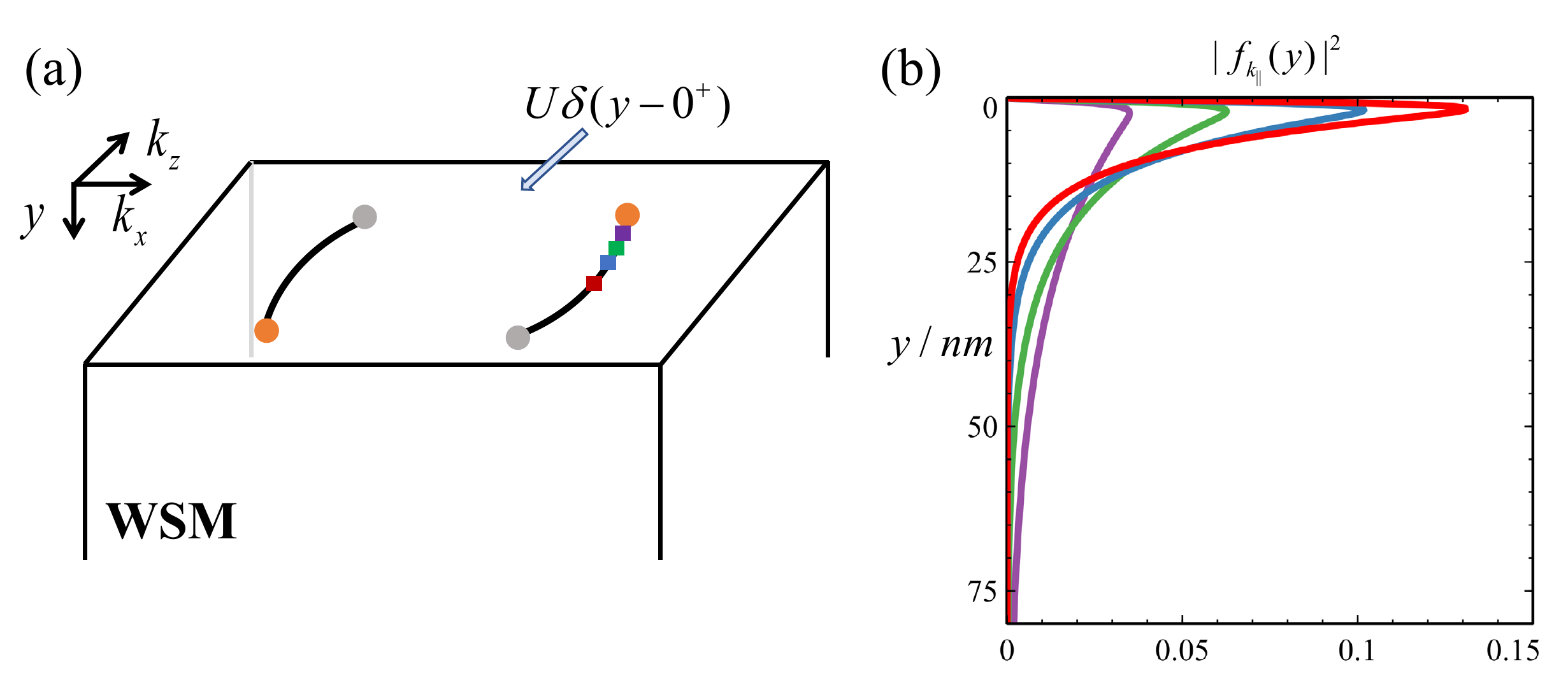}
\caption{(a) Schematic of the electrostatic potential imposed on the top surface of WSM.
(b) The dispersion of $f_{\bm{k}_\parallel}$ along $y$-direction, the color of curves is corresponding to solid squares in (a) for different $k_z$ channels, here $k_x=k_1$ .}
\label{fig2}
\end{figure}

\begin{figure*}
\center
\includegraphics[width=0.9\linewidth]{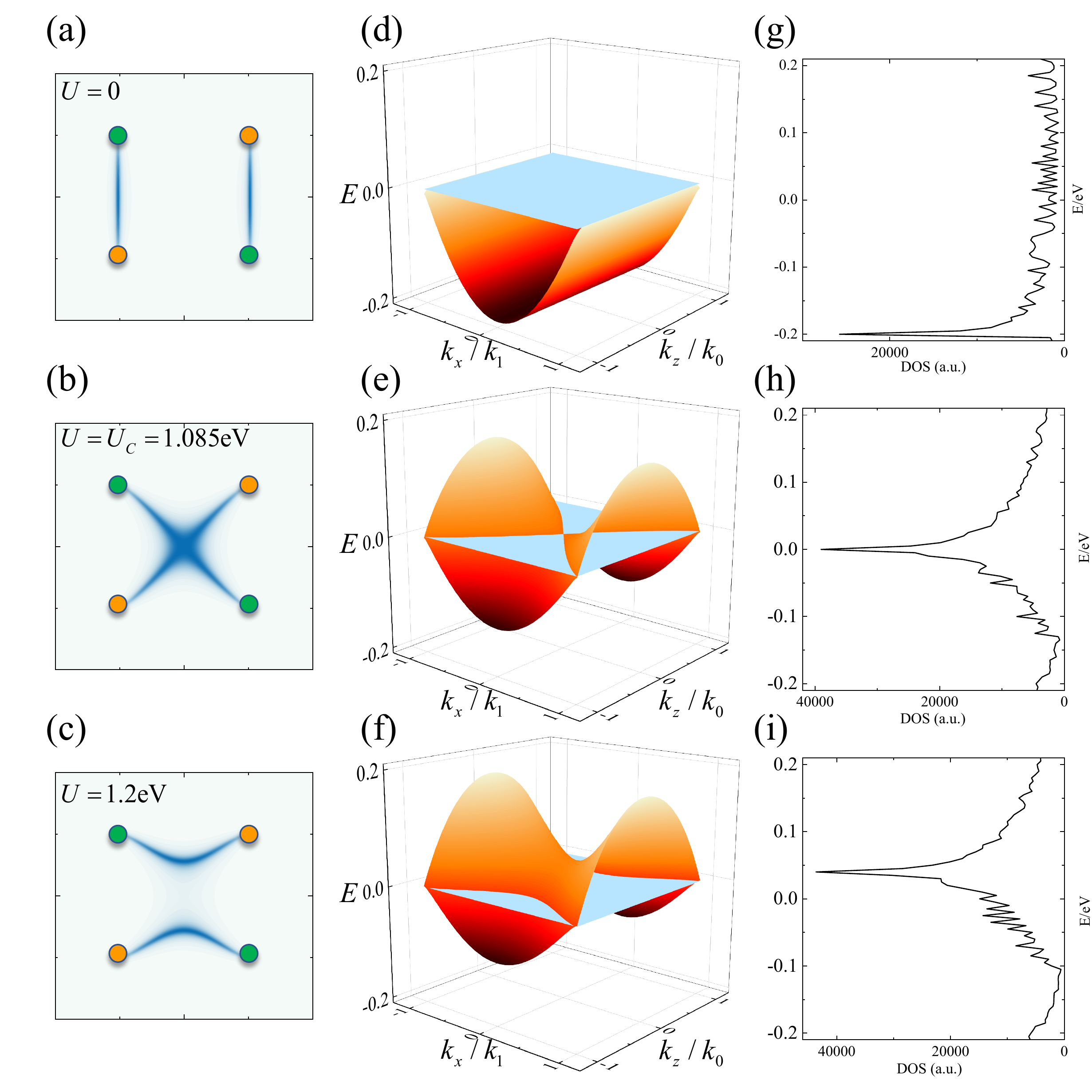}
\caption{(a)-(c)Fermi arcs spectra for different surface potential $U$.
(d)-(f) Corresponding energy dispersion of FA surface states, the blue planes represent the Fermi energy.
(g)-(i) Corresponding density of state for FA surface states.
 The calculation parameters are $a=1$ nm,
  $M_1=M_2=1.25$ eV nm$^2$, $v_y=0.66$ eV nm, $k_0=k_1=0.4$ nm$^{-1}$. }
\label{fig1}
\end{figure*}

We start with an effective model of WSM with four Weyl nodes~\cite{model}
\begin{equation}\label{H}
\begin{split}
H(\bm{k})&=M_1(k^{2}_{1}-k^{2}_{x})\sigma _x  + v_yk_{y}\sigma_{y}
+ M_2(k^{2}_{0}-k^{2}_{y}-k^{2}_{z})\sigma_z,\\
\end{split}
\end{equation}
here $v_y$ is the velocity in the $y$ direction, $k_{0,1}$ and $M_{1,2}$ are
parameters, $\sigma_{x,y,z}$ are the Pauli matrices
in the pseudo-spin space. The two bands are degenerate
at four Weyl nodes $\bm{k}_W=({\pm}k_1, 0 , {\pm}k_0)$ with two FAs spanning between them respectively [cf. Fig.~\ref{fig1}(a)].
Expanding $H(\bm{k})$ around the Weyl points yields four Weyl equations
$h(\bm{k}) =  \mp 2{M_1}{k_1}{k_x}{\sigma _x} + {v_y}k_y{\sigma _y} \mp 2{M_2}{k_0}{k_z}\sigma_z$.

We are interested in the FA surface states, which
can be solved under the open boundary condition in the
$y$ direction. Consider a semi-infinite WSM that occupies the space of
$y>0$ as shown in Fig.~\ref{fig2}(a) and make the substitution $k_y \to -i\partial_y$ in Eq.~\eqref{H}.
For a given $k_z$, Eq.~\eqref{H} reduces to a 2D system in the $x$-$y$ plane. It can be verified that
for $|k_z|<|k_0|$, such an effective 2D system possesses a nonzero Chern number
with the chiral edge state that appears at its boundary. The edge states of all
$k_z$ slices comprise the FA surface states. By solving the
eigenequation $[H(k_x,-i\partial _y, k_z)-\varepsilon_0]\psi(k_x,y,k_z)=0$
under the open boundary condition $\psi({y=0})=0$,
we obtain the
dispersion and wave function of the surface state as
\begin{equation}
\begin{split}
\varepsilon_0&= M_1(k_x^2 - k_1^2),\\
\psi (\bm{k}_{\parallel}) &= f_{k_{z}}(y){e^{i{k_x}x + i{k_z}z}}\left( {\begin{array}{*{20}{c}}
   a  \\
   b  \\
\end{array}} \right),
\end{split}
\end{equation}
with $M_2>0$, ${f_{{k_{z} }}}(y) = \eta ({e^{{\lambda _1}y}} - {e^{{\lambda _2}y}})$ the spatial distribution
function, $\eta=\frac{{2{M_2}(k_0^2 - k_z^2)}}{{1 - 4M_2^2(k_0^2 - k_z^2)}}$
the normalization coefficient, and ${\lambda _{1,2}} =- \frac{1}{{2{M_2}}} \pm \sqrt {\frac{1}{{4{M_2}^2}} + (k_z^2 - k_0^2)}$.

Here, the key point is that the wave function $\psi (\bm{k}_{\parallel})$ of the surface state
relies on the in-plane momentum $\bm{k}_\parallel=(k_x, k_z)$. Specifically, wave function for
different $\bm{k}_\parallel$ possesses unequal spatial spreading in the $y$ direction
as shown in Fig.~\ref{fig2}(b). This property
opens a new route for manipulating FA surface states
by a surface potential, which can be induced by
a surface gate voltage. Consider an electric potential
imposed on the surface of the WSM, which can be
captured by ${U_s}(y) = U\delta (y -0^+)$. It induces
an energy shift of the surface states, which can be evaluated by the overlap integral
\begin{equation}
\delta \varepsilon = \int\limits_{0}^\infty  {U\delta (y - 0^+)|{f_{\bm{k}_{\parallel}}}(y){|^2}dy}.\\
\end{equation}
The energy shift has a $k_z$ dependence and can be expanded as
$\delta\varepsilon\simeq (-Ak_z^2 + B){U}$ to the second order of $k_z$
with two constants $A, B>0$ for $M_1,M_2>0$. Physically, such
potential energy induces an effective surface dispersion and so the
full energy of the surface states becomes
\begin{equation}\label{eqdis}
\begin{split}
\tilde \varepsilon (\bm{k}_\parallel)= \varepsilon_0+\delta\varepsilon\simeq  {M_1}(k_x^2 - k_1^2) + (-Ak_z^2 + B){U}.
\end{split}
\end{equation}
It means that the band shape of the surface states and then the FAs
can be continuously tuned by the surface electric potential.

The interesting
case occurs for $U>0$, such that the coefficients before $k_x^2$ and $k_z^2$
have opposite signs. This means that the surface band bends towards opposite directions
for $k_x$ and $k_z$ and becomes a hyperbolic paraboloid; see Figs.~\ref{fig1}(e,f).
The hyperbolic paraboloid surface band contains a saddle point, which
is the key to understand the Lifshitz transition of the FAs.
Setting the Fermi energy to zero, the FAs defined by the intersection curves
between the surface band and Fermi surface undergo continuous modifications
as the surface potential $U$ increases.
The critical point for the Lifshitz transition takes place
when the saddle point of the surface band lies exactly at the Fermi energy [cf. Fig.~\ref{fig1}(e)],
where the two FAs cross each other [cf. Fig.~\ref{fig1}(b)].
As $U$ increases further, the saddle point
is lifted above the Fermi energy and accordingly,
the two FAs split again but change their way connecting the Weyl nodes [cf. Figs.~\ref{fig1}(c,f)].

In addition to the analysis based on the continuous model,
we investigate the effect of the surface potential on
the FAs by the lattice model of Eq.~\eqref{H}.
By substituting $k_{i=x,y,z}\rightarrow a^{-1}\sin k_ia$, $k_i^2\rightarrow2a^{-2}(1-\cos k_ia)$
and performing the partial Fourier transformation in the $y$ directions, we obtain
\begin{equation}\label{lat_y}
\begin{split}
 {H_{{\bm{k}_\parallel}}}(y) &= \sum\limits_{y,{k_x},{k_z}} {\psi_{{\bm{k}_\parallel}}^\dag (y)[{M_1}(\frac{2}{{{a^2}}}\cos {k_x} + k_1^2 - \frac{2}{{{a^2}}}){\sigma _x}}  \\
  &+ {M_2}(\frac{2}{{{a^2}}}\cos {k_z} + k_0^2 - \frac{4}{{{a^2}}}){\sigma _z}]{\psi_{{\bm{k}_\parallel}}}(y) \\
  &+ \sum\limits_{y,{k_x},{k_z}} {\psi_{{\bm{k}_\parallel }}^\dag (y)[\frac{{{v_y}{\sigma _y}}}{{2ai}} + \frac{{{M_2}{\sigma _z}}}{{{a^2}}}]{\psi_{{\bm{k}_\parallel }}}(y + 1)}  + \text{H.c.} \\
\end{split}
\end{equation}
with $a$ the lattice constant and ${\psi_{{\bm{k}_\parallel }}}(y) =[{\psi_{{1,\bm{k}_\parallel }}}(y), {\psi_{{2,\bm{k}_\parallel }}}(y)]$ the two-component Fermi operator.
An onsite potential $U$ is introduced to the outmost layer ($y=0$) of the lattice.
Solving the effective 1D chain in the $y$ direction under the open boundary condition
for each in-plane momentum
$\bm{k}_\parallel$ yields the surface states.
The surface bands for different electrostatic potential $U$ are plotted in Figs.~\ref{fig1}(d)-(e).
As expected, a surface potential can induce a severe deformation of the surface band,
and most importantly, a saddle point emerges. For our parameters, the critical value of $U$ for Lifshitz transition is $U_c\approx 1.085$ eV. The FAs
can be revealed by the spectral function $\mathcal{A}(E,\bm{k}_\parallel)=-(1/\pi)\text{Im}G^R(E)$
at the Fermi energy ($E=0$) with the surface Green's function defined as
\begin{equation}\label{gij}
{G^R}(E) = \sum\limits_n {\frac{{{u_{0n}}u_{0n}^*}}{{E - {E_n} + i0^+ }}},
\end{equation}
where $u_{0n}$ is the surface component ($y=0$) of the
eigenvector corresponding to the $n$-th eigenvalue $E_n$.
The configurations of FAs in Figs.~\ref{fig1}(a)-(c) are consistent
with those of the surface bands in Figs.~\ref{fig1}(d)-(f).

We have established the relation between the topological
Lifshitz transition of the FAs and the underlying
saddle points in the surface band, the latter indicating the
existence of the van Hove singularity
in the density of states (DOS) of the surface band. Therefore,
the Lifshitz transition of the FAs indicates there exists the
van Hove singularity at the Fermi energy.
In Figs.~\ref{fig1}(g-i), we plot the DOS calculated by
$\text{DOS}(E)=\sum_{\bm{k}_\parallel}\mathcal{A}(E,\bm{k}_\parallel)$.
One can see that the energy of the van Hove singularity
tracks that of the saddle point and equals the Fermi energy
at the critical point of Lifshitz transition.
Given that a variety of physical effects such as
the enhancement of electron-electron interaction and the disturbance of many-body ground states \cite{fleck1997magnetic,rice1975new,gonzalez2008kohn} are closed related
with the van Hove singularity of the band, our work opens the
possibility to realize interesting effects
in the surface states of the WSM by the surface gate induced
Lifshitz transition.

\section{Transport signatures of Fermi-arc Lifshitz transition}\label{c3}

\begin{figure}
\center
\includegraphics[width=1.0\linewidth]{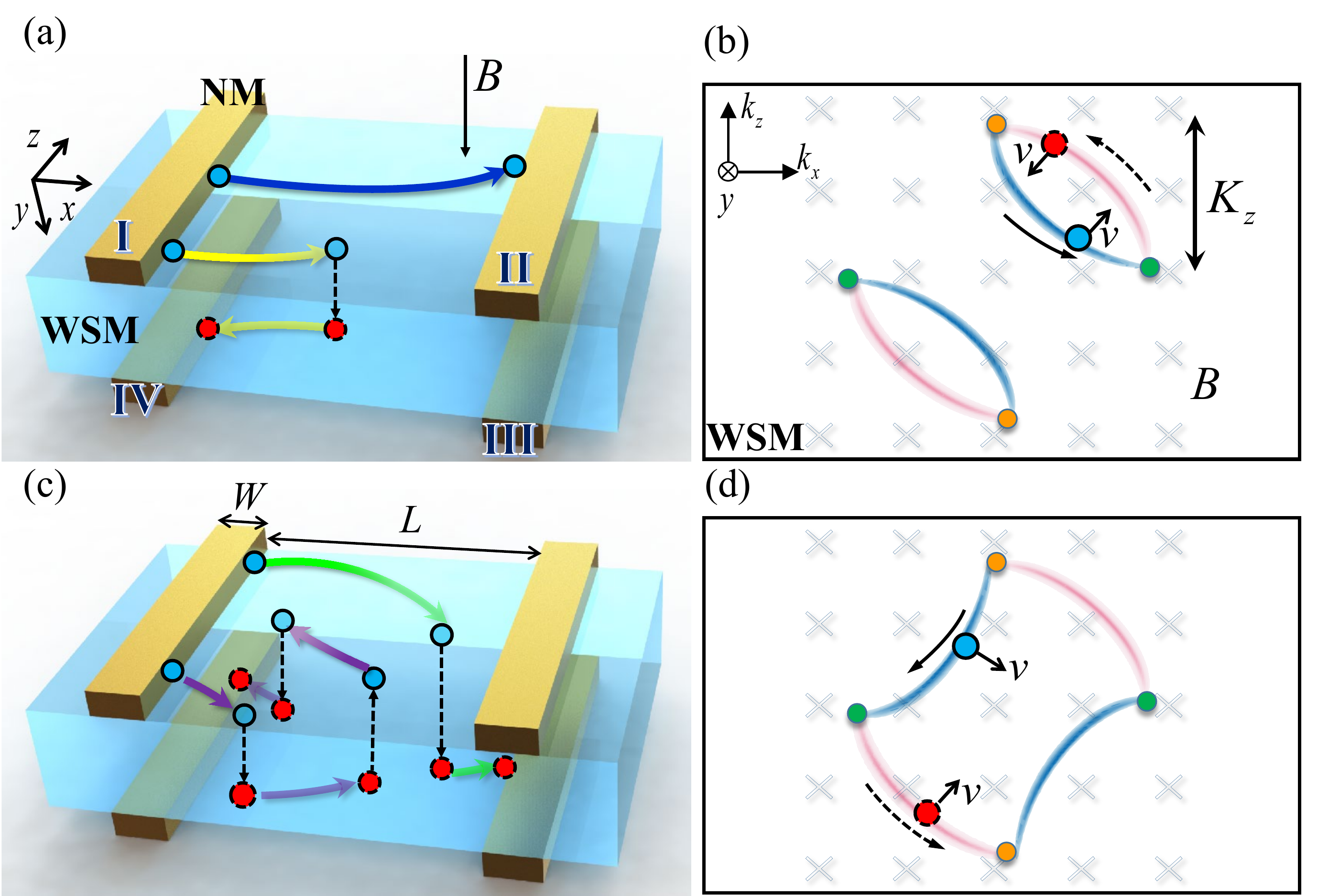}
\caption{The trajectories of electrons in real and momentum space, with the surface potential $U=0.5$ eV(a,b) and $U=1.2$ eV(c,d) at the top surface, the azimuthal angle is taken as $\theta=\pi/4$.
(a,c) Schemes of the planar junction consist of multiple normal metal (N) electrodes deposited
on top and bottom of the WSM and the scattering of particles at the surface.
The trajectories of electrons are sketched as colored arrows, and the channels provided by chiral Landau bands in the bulk is marked by dashed lines, respectively.
(b,d) The momentum of electrons slides along the FAs and transmmit between top and bottom surfaces, driven by the Lorentz force.
The curves(circles) in blue or red color represent FAs(electrons) on the top or bottom surfaces respectively.
The thickness of the WSM and the N electrodes along the $y$ direction
is 100 nm and 50 nm, respectively. The length of the coupling area between the WSM and the N electrode is $W=20$ nm,
the separation between two electrodes is $L=80$ nm.}

\label{fig3}
\end{figure}

In this section, we investigate quantum transport
in the device sketched in Fig.~\ref{fig3}(a) under a magnetic field and show that
the nonlocal conductance can provide a decisive signature of the Lifshitz transition of
the FAs. The device is fabricated by depositing multiple
strip electrodes (I-IV) on both the top and bottom
surfaces of the WSM; see Fig.~\ref{fig3}(a).
General orientations of the FAs and the stripe electrodes are considered.
Without loss of generality, we set the normal of the stripe electrodes to the $x$ direction
and perform a rotation to the WSM or the Hamiltonian~\eqref{H} by an angle $\theta$ about the $y$ axis,
or explicitly, $\tilde{H}(\bm{k})=H(U_y^{-1}\bm{k})$ with $U_y(\theta)=\left(\begin{matrix}\cos \theta & 0 & -\sin \theta \\
            0 & 1 & 0 \\
            { \sin \theta } & 0 & {\cos \theta }\end{matrix}\right)$.
Accordingly, the configurations of the FAs shown in Fig.~\ref{fig1}
undergo the same rotation; see Figs.~\ref{fig3}(b,d). The stripe electrodes
are just normal metals which can be described by the effective Hamiltonian
as $H_N(\bm{k})=C\bm{k}^2-\mu_N$
with $C$ the parameter determined by the effective mass
and $\mu_N$ the chemical potential.
We here focus on the surface transport with a negligible density
of the bulk states and so set the chemical potential of the WSM
to zero for simplicity.

\begin{figure*}
\center
\includegraphics[width=1.0\linewidth]{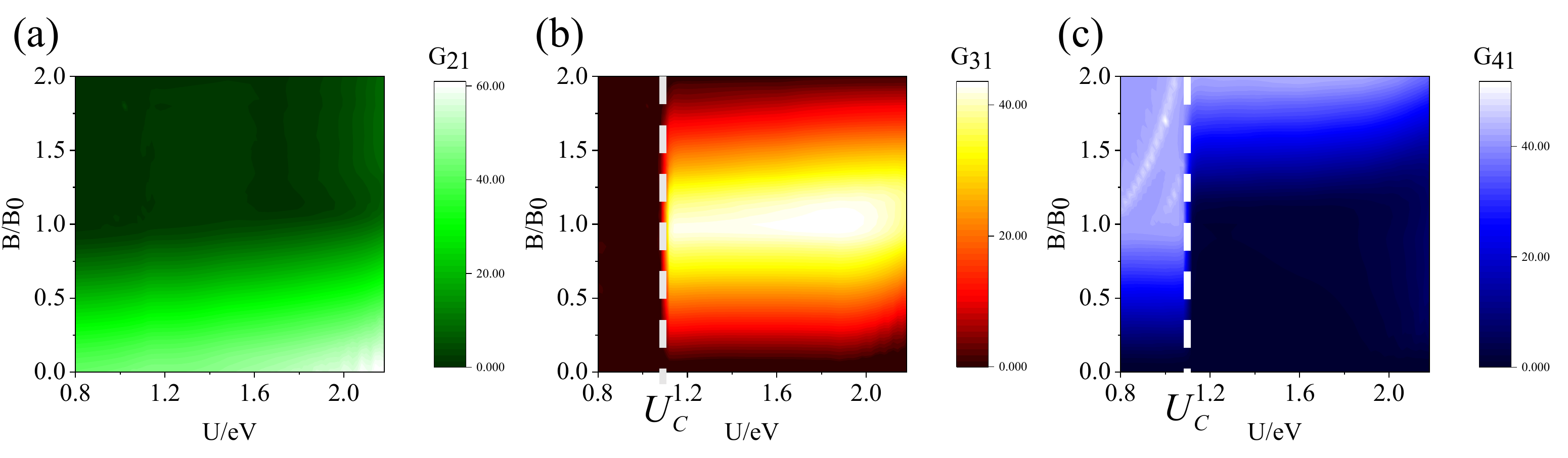}
\caption{The nonlocal differential conductance $G_{ij}$ between (a) electrodes I and II (b) electrodes I and III (c) electrodes I and IV for different  magnetic fields and surface potential. Here $C=1$ eV, $\mu_N=2.2$ eV, $V_1=5$ meV, other parameters are the same as those in Fig.~\ref{fig1}. }
\label{fig4}
\end{figure*}

Consider an electron injected from electrode I into the surface of the
WSM. Without a magnetic field, it propagates straightforwardly into electrode II.
Interesting situations take place when a magnetic field $B$ is imposed
in the $y$ direction. Due to the Lorentz force, electrons are driven to
slide along the FAs; see pictures in
the momentum space [Figs.~\ref{fig3}(b,d)] and
the corresponding curved trajectories in the real space [Fig.~\ref{fig3}(a,c)].
Because the FAs are terminated at the Weyl nodes, a closed loop
for electron transport is composed of the FAs on both surfaces and the
chiral Landau bands inside the bulk, so-called the
Weyl orbit~\cite{potter2014quantum}. As a result,
the trajectory of an incident electron
is completely determined by the configuration of the Weyl orbit.
As discussed in the previous section, a surface gate deposited on
the top surface can modify the FAs therein, while
the chiral Landau band and the FAs on the bottom surface are unaffected.
A direct result is that the Weyl orbit possesses distinctive
configurations and topology
before and after the Lifshitz transition of the top FAs;
compare Fig.~\ref{fig3}(b) and Fig.~\ref{fig3}(d).

For the Weyl orbit in Fig.~\ref{fig3}(b),
electrons injected from electrode I may have two kinds of trajectories.
For the first one, electrons propagate only on the top
surface until they enter electrode II; see the blue arrowed
line in Fig.~\ref{fig3}(a). In the second case,
electrons first slide along the FAs on the top surface
and reach the Weyl node,
then transfer through the chiral Landau band to the bottom surface,
propagate backward and finally reach electrode IV;
see the yellow arrowed lines in Fig.~\ref{fig3}(a).
Based on this picture, one can infer that
as electrode I is biased, the injected electrons can
only contribute to current flowing in electrode II and IV,
because no electron can reach electrode III.
The distribution of the current in II and IV relies
on both the magnetic field and the distance $L$ between electrodes I and II.

Once the topological Lifshitz transition takes place
for the FAs on the top surface of the WSM, the two isolated
Weyl orbits in Fig.~\ref{fig3}(b) merge into a single but larger one
as shown in Fig.~\ref{fig3}(d). The electrons
injected from electrode I change their trajectories
accordingly. Apart from the direct propagation from electrode I to II,
electrons can also transfer to the bottom surface and reach electrode III;
see the green arrowed lines in Fig.~\ref{fig3}(c). The main difference
between this regime and that in Fig.~\ref{fig3}(a) is that
the electrons transferred to the bottom surface do not reverse
their velocity in the $x$ direction. It can be expected that
the nonlocal conductance between electrode I and III will exhibit
a switch-on effect during the Lifshitz transition of the FAs, which
provides a decisive signal for its detection. Moreover, once the magnetic field exceeds
a critical valve, the electrons can propagate along a more complicated trajectory and enter electrode IV;
see the purple arrowed line in Fig.~\ref{fig3}(c).

Next, we perform numerical calculation of the quantum transport
in the device sketched in Fig.~\ref{fig3}(a) to verify the
semiclassical picture discussed above. The calculation is conducted
by discretizing the effective Hamiltonian
$\tilde{H}(\bm{k})$ and $H_N(\bm{k})$ on a cubic lattice. It is assumed that the size of the strip electrodes in the
$z$ direction is much larger than the Fermi wavelength
and their contacts with WSM assure the conservation of the momentum $k_z$.
Then by taking $k_z$ as a parameter,
the 3D system can be decomposed into
a series of 2D slices labeled by $k_z$,
which effectively accelerates the numerical calculation.
For the magnetic field effect, the Landau gauge
$\bm{A}=({0, 0, -B x})$ is adopted
so that the Peierls substitution
$\bm{k}\rightarrow -i\bm{\nabla}\pm e\bm{A}/\hbar$ (taking $e>0$)
retains the conservation of $k_z$.

The nonlocal differential conductance
$G_{ij}=\partial I_i/\partial V_j$ at zero temperature can be calculated using
the Landauer-B\"{u}ttiker formula by summing up the contributions of all the $k_z$ channels as
\begin{equation}\label{G}
G_{ij}(eV)= \frac{{e^2}}{h}\sum_{k_z}\text{T}_{ij}^{k_z} ,
\end{equation}
where $\text{T}_{ij}^{k_z} $ is the transmission probability of the $k_z$ slice
from electrode $j$ to $i$ calculated using KWANT \cite{groth2014kwant}.

Consider a nonzero bias $V_1$ is applied to electrode I, and the nonlocal conductances
between electrode I and the other three as a function of the surface potential
$U$ and magnetic field $B$ are plotted in Fig.~\ref{fig4}.
The unit of the magnetic field is set to the critical value $B_0=\hbar K_z / (eL)$ with
$K_z$ the span of the FA in the $k_z$ direction [Fig.~\ref{fig3}(d)] and $L$ the separation between
electrodes I and II [Fig.~\ref{fig3}(c)]. $B_0$ is the critical field strength to drive all
incident electrons to the Weyl node and penetrate into the bulk before they reach
electrode II. As discussed previously, a surface gate potential $U$ can induce
the Lifshitz transition in the FAs on the top surface. We denote the gate voltage corresponding to
the transition point by $U_c$. Rich information is involved in the
conductance patterns in Fig.~\ref{fig4}, which can be interpreted
by looking at the $B$-dependence of the conductances before and after
the Lifshitz transition.

For $U<U_c$, since the FAs on the top and bottom surfaces form two independent Weyl orbits,
the right-moving electrons from electrode I can only slide along the right
loop in Fig.~\ref{fig3}(b). Moreover, the states of the FAs on the top and bottom
surfaces possess opposite group velocities. As a result, electrons have two kinds of
trajectories, propagating from electrode I to II directly or,
transmitting into the bottom surface and being reflected back into electrode IV [cf.
Figs.~\ref{fig3}(a)]. As $B$ increases from zero, more incident electrons
takes the latter trajectory [yellow arrowed lines in Fig.~\ref{fig3}(a)].
As a result, $G_{41}$ increases while $G_{21}$ decreases for a larger $B$ as shown in Figs.~\ref{fig4}(a,c).
Meanwhile, no electron can reach electrode III and so the nonlocal conductance $G_{31}$ vanishes
in this regime; see Fig.~\ref{fig4}(b).

For $U>U_c$, the Lifshitz transition occurs which gives rise to a drastic modification
of the trajectories of electrons and accordingly, different $B$-dependence of the conductances.
First, $G_{21}$ always decreases for a stronger magnetic field, which is similar
to the situation for $U<U_c$. The main difference is the $B$-dependence of $G_{31}$ and $G_{41}$.
Due to the unique trajectory sketched by the green arrowed lines in Fig.~\ref{fig3}(c),
$G_{31}$ is switched on by the Lifshitz transition and increases with $B$ for $B<B_0$; see Fig.~\ref{fig4}(b).
Accordingly, $G_{41}$ is switched off in this region [Fig.~\ref{fig4}(c)] because the yellow trajectory in Fig.~\ref{fig3}(a)
is absent. Instead, the electrons need to take more complex trajectory sketched by the purple arrowed lines
in Fig.~\ref{fig3}(c) to reach electrode IV, which needs a stronger magnetic field.
Such abrupt changes of $G_{31}$ and $G_{41}$ with $U$ manifest the
topological Lifshitz transition of the FAs and the Weyl orbits.
As $B$ reach the critical value $B_0$, all electrons penetrate to the bottom
surface and lead to a maximum $G_{31}$; see Fig.~\ref{fig4}(b).
As $B$ increases further, the purple trajectory in Fig.~\ref{fig3}(c) comes into play,
which results in an increase of $G_{41}$ accompanied by a decrease of $G_{31}$;
see Figs.~\ref{fig4}(c) and ~\ref{fig4}(b).

%In conclude, the magnetic field can drive the momentum of surface electrons sliding along the so-called Weyl orbitals between top and bottom surface, and the Lifhitz transition can merge two Weyl orbitals into a single one, providing the scattering channel along $+x$ direction for electrons at bottom surface, which can finally contribute to the current at electrode III.
%Therefore, for $U>U_c$ a significant conductance between electrodes I and III can be measured as long as the magnetic field is within a appropriate range($0<B<2B_c$), which could not occur at $U<U_c$ and can serve as the decisive criterion of the topological Lifshitz transition of FAs.
%The conductance spectrum has a clear demarcation between the two regions at $U=U_c$, which is coincident exactly with the two topologically inequivalent phases.
%The difference response of conductance spectrum to the magnetic fields before and after phase transition
%stems from the high anisotropy of FA configurations as well as their combination with chiral Landau band in the bulk,
%which may also be utilized to explore other chiral magnetic effects in WSM.
%the income electrons may arrive at electrode II straightly, pass through to the bottom surface and be detected at electrode III, or even cross back to the top surface and return to electrode I.

\begin{figure}
\center
\includegraphics[width=1.0\linewidth]{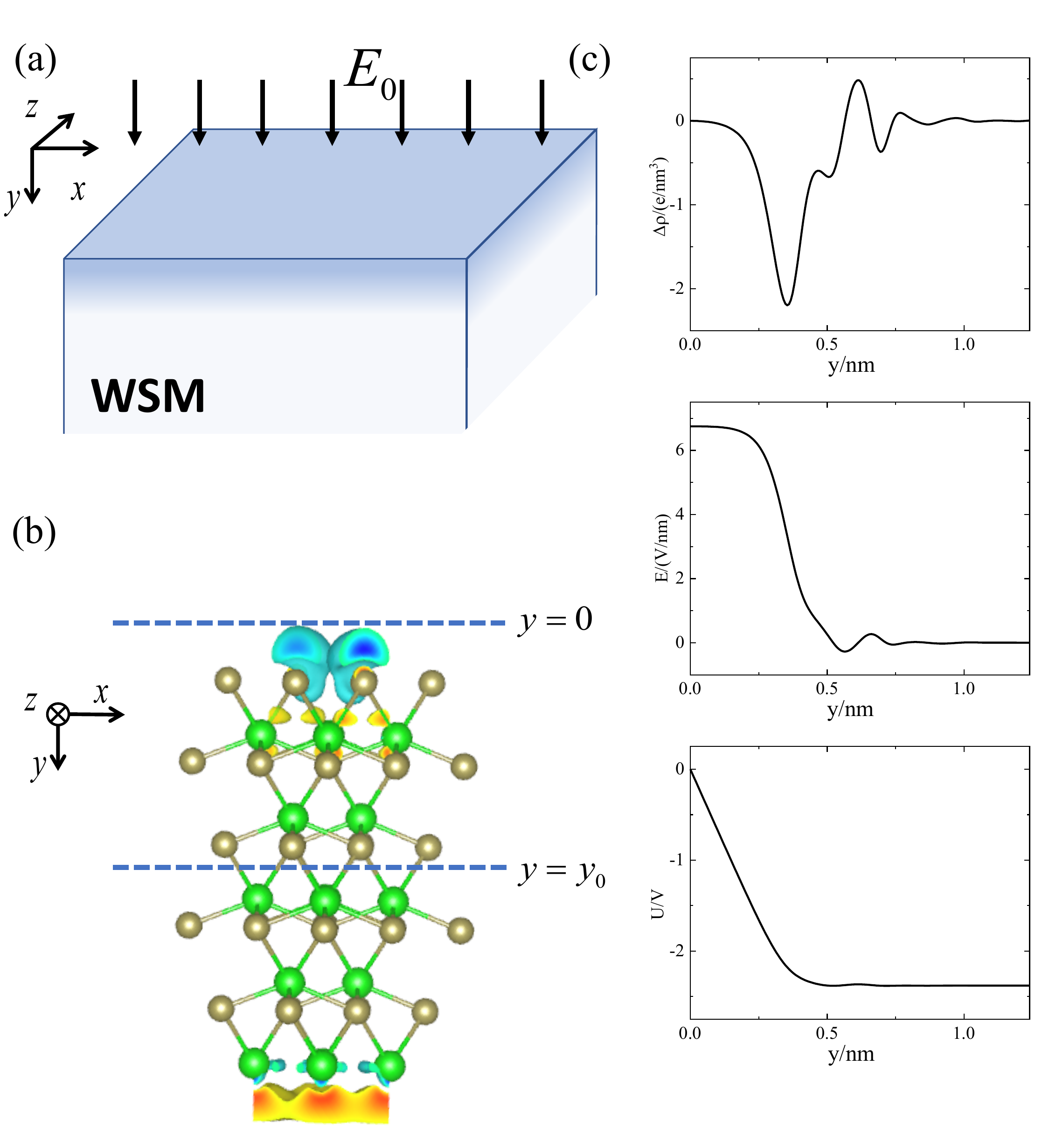}
\caption{(a) Scheme of the potential resulting from the difference of charge density near the surface of WSM, the external electric field $E_0$ is imposed perpendicular to the open surface of ZrTe.
(b) The difference of charge density between $E_0=5$V/nm and $E_0=0$ condition, the yellow and blue parts near the surface represent the increase and decrease of charge density respectively, the zero electric field reference surface is taken as $y_0\approx 1.24$ nm.
(c) The calculation of surface electrostatic potential induced by the external electric field. The top, middle and bottom panels respectively depict the distribution of charge density difference, electric field intensity and electric potential along the $y$ direction.}
\label{fig5}
\end{figure}

\section{Discussions}\label{c4}
We discuss the experimental implementation of our proposal.
In the previous discussion, all results are obtained based on the
minimal model~\eqref{H} of the WSM and the effect of the surface
gate is introduced phenomenologically by the potential $U$.
From Fig.~\ref{fig2}, one can see that the Lifshitz transition of FAs
can be driven by a surface potential with $U_c\sim 1$ eV. In the following,
we show that such a surface potential can be achieved in real WSMs.
In reality, the establishment of the surface potential
by the gate voltage is sketched in Fig.~\ref{fig5}(a).
A surface gate creates a vertical electric field, which drives
free charges to the sample boundary,
where these charges rearrange and achieve equilibrium again
due to the screening effect. As a result, a finite ingredient of the charge density
and thus that of the electric potential are induced by the surface gate,
which leads to a potential difference between the surface and bulk electrons. This
is the physical origin of the parameter $U$ in the previous calculation.

We show that the strength of the surface potential $U_c\sim1$ eV
required by the Lifshitz transition of the FAs can be realized
in real WSMs. We here take a typical WSM, ZrTe~\cite{weng2016coexistence} as an example,
and perform first-principles calculations on the charge density difference induced by an external electric field
using VASP software package~\cite{kresse1996efficiency,kresse1996efficient}.
The calculation is performed on a three-layer ZrTe slab with a $1.4 $ nm vacuum,
and an external electric field $E_0$ is applied perpendicular to the open surface, which
is created by a surface gate in the experiment.
The sampling of the Brillouin zone in the self-consistent process is taken as the grid of
$ 18\times 18 \times 13$, and the exchange-correlation potential is
treated within the generalized gradient approximation~\cite{kohn1965self} of the Perdew-Burke-Ernzerhof type~\cite{perdew1996generalized}.
The difference of the charge density distribution between $E_0=5$~V/nm and $E_0=0$ is plotted in Fig.~\ref{fig5}(b).
As is shown, negative and positive charge density difference is accumulated on the top and bottom boundaries, respectively.
The fluctuation of the charge density on the atomic scale in the $x$-$z$ plane is not important
so that we take its average value $\Delta \rho$ in this plane for simplicity and focus on its distribution
in the $y$ direction. The charge distribution induced by the external electric field
is plotted in Fig.~\ref{fig5}(c).Using $\partial E/\partial y=\Delta\rho(y)/\varepsilon_0$ and $\partial U/\partial y=-E$
and taking into account the vanishing field inside the bulk, $E(y=y_0)=0$, the distribution of $E$ and $U$ in the $y$ direction can be obtained; see Fig.~\ref{fig5}(c). One can see that for an external electric field $E_0=5$~V/nm,
a considerable surface potential $U\sim 2$~eV
can be induced within the range of 0.3~nm near the surface. Such a result proves
that a surface potential sufficient to drive the Lifshitz transition of FAs
can be induced by the surface gate voltage, showing the feasibility of our proposal.

The minimal WSM model in Eq.~\eqref{H} contains a single pair of the FAs on each surface,
which has been reported in NbIrTe$_4$ (TaIrTe$_4$)~\cite{Koepernik16prb,belopolski2017signatures,Haubold17prb,Ekahana20prb}, WP$_2$~\cite{Yao19prl},
MoTe$_2$~\cite{Wang16prl}, and YbMnBi$_2$~\cite{borisenko2019time}.
In other WSMs however, more than one pair of FAs exist. We argue that
our main conclusion of the gate tunable Lifshitz transition of the FAs should maintain
in all these realistic cases. First, the feature of the surface bands that can sustain
FAs generally allow the existence of the saddle points, which is the key scenario
of the Lifshitz transition; Second, the surface bands can be generally tuned by
the surface gate voltage, which is the other ingredient of our scheme. Moreover,
for the surface bands in realistic materials, some saddle points may
lie naturally near the Fermi energy
so that the surface potential required by the Lifshitz transition
may be much smaller than $1$~eV in our model,
which further facilitates its implementation.
As for the detecting proposal, the stripe electrodes
can be fabricated on the WSM by state-of-the-art techniques~\cite{li2020fermi,chen2018finite,ghatak2018anomalous}.
FAs with regular shapes are profitable to our proposal, as the spatial trajectories of electrons
possess regular configurations accordingly. Moreover, a big separation
between Weyl points in the momentum space~\cite{Sun15prb,Koepernik16prb,
Change16sa,Yao19prl,borisenko2019time,sie2019ultrafast} is also beneficial for our proposal,
in which visible transport signatures can be expected.

%In our transport calculations, for simplicity, zero chemical potential
%was taken in the WSM, where there is a vanishing density of bulk states.
%In real materials with finite density of bulk states,
%our main results remain unchanged as long as the presence of bulk states will only cause
%certain conductance leakage of surface electrons, but does not change the current qualitative results.

\section{Summary}\label{c5}
To conclude, we show that topological Lifshitz transition of FAs can be induced by
a surface gate voltage. Such a transition is attributed to the existence of saddle points
in the surface bands, which meets the Fermi energy at the critical transition point.
A direct result due to the Lifshitz transition is the abrupt change of the magnetic Weyl orbits
composed of both the FAs and bulk Landau bands. Such an effect can be detected
by nonlocal transport signatures and the Lifshitz transition can be visibly revealed by
the switch on and off of the nonlocal conductances.

\begin{acknowledgements}
This work was supported by the National Natural Science Foundation of
China under Grant No. 12074172 (W.C.), No.  12222406  (W.C.) and
No. 12174182 (D.Y.X.), the State Key Program for Basic
Researches of China under Grants No. 2021YFA1400403 (D.Y.X.), the Fundamental Research
Funds for the Central Universities (W.C.), the startup
grant at Nanjing University (W.C.) and the Excellent
Programme at Nanjing University.
\end{acknowledgements}

\end{document}